\documentclass[a4paper,10pt,notitlepage,fleqn]{article}
\RequirePackage{latexsym}
\RequirePackage[english]{babel}
\RequirePackage{amssymb}
\RequirePackage{amsbsy}
\RequirePackage{amsmath}
\usepackage[T1]{fontenc}
\usepackage[dvips]{graphicx}
\usepackage[dvips]{color}
\newcommand{\beq}{\begin{eqnarray}}
\newcommand{\eeq}{\end{eqnarray}}
\newcommand{\ud}{\mathrm{d}}

\newcommand{\lt}{\left(}
\newcommand{\rt}{\right)}
\newcommand{\lqu}{\left[}
\newcommand{\rqu}{\right]}

\newcommand{\dota}{\dot{a}}
\newcommand{\ddota}{\ddot{a}}

\newcommand{\pop}{\hat{\phi}}

\newcommand{\pa}{P_a}

\newcommand{\ptau}{\partial_\tau}
\newcommand{\para}{\partial_a}

\newcommand{\lp}{\ell_{\rm p}}
\newcommand{\hub}{\mathcal{H}}

\newcommand{\bra}[1]{\mbox{$\langle #1\! \mid$}}

\newcommand{\ket}[1]{\mbox{$\mid \!#1\rangle$}}

\newcommand{\vm}[1]{\mbox{$\langle #1\rangle$}}
\newcommand{\hh}{\hat{H}_\phi}

\newcommand{\kcs}{\ket{\chi_{\rm s}}}
\newcommand{\kck}{\ket{\chi_k}}
\title{\bf A radiation-like era before inflation}
\author{
Corrado Appignani$^{a}$\thanks{appignani@bo.infn.it}$\ $
and
Roberto Casadio$^{a}$\thanks{casadio@bo.infn.it}
\\
\null
\\
$^a${Dipartimento di Fisica and I.N.F.N., Sezione di Bologna,}
\\
{via Irnerio 46, 40126 Bologna, Italy}
}
\begin{document}
\maketitle
\begin{abstract}
We show that the semiclassical approximation to the Wheeler-DeWitt equation
for the minisuperspace of a minimally coupled scalar field in the spatially flat de~Sitter
Universe prompts the existence of an initial power-law evolution driven by non-adiabatic
terms from the gravitational wavefunction which act like radiation.
This simple model hence describes the onset of inflation from a previous radiation-like
expansion during which the cosmological constant is already present but subleading.
\end{abstract}
\section{Introduction}
\setcounter{equation}{0}
The inflationary paradigm is a basic ingredient of the standard model of cosmology
which is well-known for solving several problems of the old big bang theory~\cite{infla}.
Many different mechanisms have been proposed to describe both the onset of such a
phase and how it ends, most of which rely on the existence of an ``inflaton'' (scalar) field.
The stress tensor of the latter plays the role of an effective cosmological constant and
drives the accelerated expansion of the universe through the Einstein equations,
but its nature -- whether fundamental or effective --  is not clearly understood yet.
It is therefore worth considering alternative mechanisms to realise such a scenario.
\par
We shall here address this problem by employing the Born-Oppenheimer approach
in the semiclassical approximation~\cite{bv} for the Wheeler-DeWitt equation~\cite{WDW}
in the minisuperspace of the cosmic scale factor $a$ and one mode of a minimally coupled
scalar field $\phi$, the latter representing matter and not the inflaton.
Our approach closely follows that of Refs.~\cite{bv,fvv} and leads to two equations
governing, respectively, the gravitational wavefunction $\psi=\psi(a)$ and the matter
quantum state $\chi=\chi(\phi)$.
It was already shown in Ref.~\cite{xi0} that quantum gravitational fluctuations become
negligible for ``large'' $a$ (late times) and, if a cosmological constant $\Lambda$ is
present, the de~Sitter evolution (that is, eternal inflation) is recovered.  
The opposite regime of ``small'' $a$ (early times) was investigated in
Ref.~\cite{asyfree}, in which we showed that quantum gravitational fluctuations
are again negligible and matter states $\chi$ become asymptotically free on approaching
the limit of the semiclassical regime.
\par
Although the consistency of such a matter behaviour with all the approximations
employed was checked {\em a posteriori\/}, possible corrections to the evolution of $a$
were not considered in Ref.~\cite{asyfree}.
In the following, we shall therefore analyse in details the equation governing
$a$ for asymptotically free matter states and show that the terms coming from the
WKB approximation for the gravitational wavefunction $\psi$, which would vanish in the
adiabatic approximation, actually dominate the evolution of the scale
factor at very early times (like the cases treated in Ref.~\cite{gf}).
In particular, we shall uniquely identify one consistent coupled dynamics for matter and
gravity in which $a$ evolves like in a radiation-dominated universe, although no radiation
is actually present.
In this scenario, the onset of inflation occurs at the end of the semiclassical regime
without the need of a dynamical inflaton field.
\par
In Section~\ref{msc}, we briefly review the general formalism of Ref.~\cite{xi0} and
the relevant results from Ref.~\cite{asyfree}.
The equation for gravity will then be analysed in the very early regime in
Section~\ref{Sgeq} and the equation for matter in Section~\ref{Smeq}, where
we uniquely identify the radiation-like behaviour for $a$.
We finally comment and compare with previous results in Section~\ref{conc}.
\par
We use units with $\hbar=c=1$.
\section{Minisuperspace semiclassical cosmology}
\label{msc}
\setcounter{equation}{0}
The Wheeler-DeWitt equation in the minisuperspace of
one mode of wavenumber $k$ of a minimally coupled scalar field $\phi=\phi(\tau)$
and the scale factor $a=a(\tau)$, can be written as
\beq
\hat H\,\Psi(a,\phi)=\left(\hat H_G+\hat H_\phi\right)\Psi(a,\phi)=0
\ ,
\label{wdw}
\eeq
where $H_G$ and $H_\phi$ are the Hamiltonians for $a$ and $\phi$
respectively~\cite{bv,fvv,xi0}.
From the Born-Oppenheimer decomposition of the universe wavefunction,
\beq
\Psi(a,\phi)=\psi(a)\,\chi(\phi;a)
\ ,
\eeq
after several manipulations and redefinitions,
one obtains an equation for the gravitational part
(see Eq.~(14b) of Ref.~\cite{xi0} with tildes omitted for simplicity),
\beq
\lt \frac{\lp^2}{2\,a}\,\para^2 + \frac{a^3\,\Lambda}{6\,\lp^2}
+\vm{\hat H_\phi}\rt\psi
=-\frac{\lp^2}{2\,a}\,\bra{\chi_k}{\para^2}\ket{\chi_k}\,
\psi
\ ,
\label{grav}
\eeq
where $\Lambda$ is the cosmological constant,
and one for the matter part (see Eq.~(14a) of Ref.~\cite{xi0}),
\beq
\left[\frac{\lp^2}{a}\frac{\partial_a\psi}{\psi}\,\partial_a
+\left(\hat H_\phi-\vm{\hat H_\phi}\right)\right]\ket{\chi_k}
=-\frac{\lp^2}{2\,a}\left(\partial_a^2-\bra{\chi_k}{\partial_a^2}\ket{\chi_k}\right)
\ket{\chi_k}
\ ,
\label{ma}
\eeq
with
\beq
H_\phi=\frac{1}{2}\left(\frac{\pi_\phi^2}{a^3}+a\,k^2\,\phi^2\right)
\ .
\eeq
It is important to remark that no approximation was invoked this far and
Eqs.~\eqref{grav} and~\eqref{ma} are therefore equivalent to Eq.~\eqref{wdw}.
\par
From now on, we require that $a$ behaves like a (semi)classical quantity by
assuming the validity of the WKB~approximation for the gravitational wavefunction,
\beq
\psi
\simeq
\frac{1}{\sqrt{\pa}}\,e^{-i\,\int^a \pa(x)\,\ud x}
\ ,
\label{wkb}
\eeq
where 
\beq
\pa = -\lp^{-2}\,a\,\dota
\label{Pa}
\eeq
is the classical momentum associated to $a$.
Again, after some manipulations and redefinitions, this yields the approximate matter
equation (see Eq.~(18) of Ref.~\cite{xi0})
\beq
\lqu
1-
\frac{3\,i\,\lp^2}{2\,a^3\,\hub}
\lt 1 + \frac{2\,\dot\hub}{3\,\hub^2}\rt
\rqu
\lt i\,\ptau - \hh \rt \kcs
\simeq
\frac{\lp^2}{2\,a^3\,\hub}
\,\Delta\hat O\,\kcs
\ ,
\label{mat}
\eeq
where $\hub\equiv\dot a/a\equiv \partial_\tau a/a$ is the Hubble ``parameter''
and the cosmic time $\tau$ is introduced by means of the semiclassical (WKB)
relation~\cite{bv}
\beq
\lp^2\lt \partial_a\log\psi\rt \partial_a\simeq i\,a\,\partial_\tau
\ .
\label{tau}
\eeq
Note that the matter state appears in Eqs.~\eqref{grav} and \eqref{mat} in
different representations, namely
\beq
\kcs = \exp\lt -i \int^\tau \!\! \vm{\hh}\, \ud t \rt
\kck
\ ,
\label{rela}
\eeq
but the expectation value of the matter Hamiltonian is the same for both,
\beq
\vm{\hh}\equiv\bra{\chi_k}\hh\kck =\bra{\chi_{\rm s}}\hh\kcs
\ .
\eeq
The terms in the right hand sides of Eqs.~\eqref{grav} and \eqref{mat}
represent quantum gravitational fluctuations generated by the presence of matter.
In particular,
\beq
\hat O=\frac{2\,i}{\hub}\,\vm{\hh}\,\ptau
+\frac{1}{\hub}\,\ptau^2
+3\,i\lt 1 + \frac{2\,\dot\hub}{3\,\hub^2}\rt
\hh
\label{rhsm}
\eeq
and $\Delta \hat{O} \equiv \hat{O} - \bra{\chi_{\rm s}}{\hat{O}}\kcs$.
\par
In Ref.~\cite{asyfree}, we considered the above equations for the particular case
of the spatially flat de~Sitter universe,
\beq
a_{\rm dS}=\alpha\,e^{\hub_0\,\tau}
\ ,
\label{ds}
\eeq
where $\hub=\hub_0=\sqrt{\Lambda/3}$ and $\alpha$ are constants.
We were then able to show that the quantum gravitational fluctuations in the matter
equation~\eqref{mat} determined by the operator~\eqref{rhsm} become negligible
and the scalar field appears asymptotically free at very early times,
when $a_{\rm dS}\sim k\,\lp$ (which was named regime~I in Ref.~\cite{xi0}).
In fact, for
\beq
y=\frac{\lp^2}{2\,\hub\,a^3} \gg 1
\ .
\label{ydef}
\eeq
it is possible to write the scalar field wavefunction as
\beq
\kcs\sim
\ket{\chi_{\rm asy}}
=
\sqrt{\mathcal{N}}\,\exp\lt -i\,\int \frac{3\,\lp^2\,\kappa^2}{2\,a^3}\, \ud t\rt
\ket{\kappa}
\ ,
\label{wematt}
\eeq
where $\ket{\kappa}$ are orthogonal eigenstates of the scalar
field momentum,
\beq
\frac{\hat\pi}{\sqrt{3}\,\lp}\ket{\kappa}=\kappa\,\ket{\kappa}
\ ,
\eeq
and $\mathcal{N}$ a normalization factor.
The corresponding matter energy (after subtracting the usual zero-point contribution)
is given by
\beq
\bra{\chi_{\rm asy}}{\hh}\ket{\chi_{\rm asy}}
=\frac{3\,\lp^2\,\kappa^2}{2\,a^3}
\simeq
\frac{k}{a}\,\vm{n}
\ ,
\label{hasy}
\eeq
where $\vm{n}$ is the mean occupation number (in terms of the usual particle
interpretation) for the mode $k$, which was estimated to be very close to zero~\cite{asyfree}.
The compatibility of these matter states with the de~Sitter evolution was checked
by showing that their energy~\eqref{hasy} remains subleading
with respect to the cosmological constant $\Lambda$ and the corresponding
quantum gravitational fluctuations also vanish in the semiclassical Friedmann
equation~\eqref{grav}.
\par
The coupled dynamics of this matter-gravity system in the asymptotic regime
was not fully exploited in Ref.~\cite{asyfree} and we shall here complete its analysis.
\section{The equation for gravity at very early times}
\label{Sgeq}
\setcounter{equation}{0}
We now want to study Eq.~\eqref{grav} for the gravitational degree of freedom
in the very early universe, back to the limit of validity of the semiclassical
approximation, that is for
\beq
a \sim a_0\equiv a(\tau_0)=k\,\lp
\ ,
\eeq
without assuming a specific time dependence
for $a=a(\tau)$~\footnote{In Ref.~\cite{asyfree},
the asymptotic regime was taken at $\tau=\tau_{\rm asy}\to-\infty$,
whereas in the present work we rescale the constant $\alpha$ in Eq.~\eqref{ds} so that
$\tau_{\rm asy}\to\tau_0>0$.
This choice will appear necessary to ensure that the scale factor grows for increasing
time (see Eqs.~\eqref{solgrav1_4} and~\eqref{solgrav1_2}).}.
\par
The expectation values in Eq.~\eqref{grav} are taken on the asymptotic matter states
given in Eq.~\eqref{wematt}.
From Eq.~\eqref{rela}, we therefore find the asymptotic behaviour
\beq
\ket{\chi_k}
\simeq
\sqrt{\mathcal{N}}\,\ket{\kappa}=
\sqrt{\mathcal{N}\,\frac{\sqrt{3}\,\lp}{2\,\pi}}\:
e^{i\,\sqrt{3}\,\kappa\,\lp\,\phi}
\eeq
which is clearly independent from $a$ and leads to asymptotically
vanishing gravitational perturbations,
\beq
\bra{\chi_k}{\para^2}\ket{\chi_k}\simeq 0
\ .
\eeq
In the asymptotic regime $y\gg 1$ (see Eq.~\eqref{ydef}) and $a\sim k\,\lp$,
the gravitational equation hence simplifies to
\beq
\lt \frac{\lp^2}{2\,a}\,\para^2 + \frac{a^3 \Lambda}{6\,\lp^2}
+\vm{\hh}\rt \psi
\simeq 0
\ .
\label{grav2}
\eeq
\par
On substituting the WKB (semiclassical) form~\eqref{wkb} for the wavefunction
$\psi=\psi(a)$, one now gets the following evolution equation for the scale factor
\beq
\frac{\lp^2}{2\,a}
\lqu \frac{3}{4\,a^2} + \frac{\ddota}{2\,a\,\dota^2}
+ \frac{5\,\ddota^2}{4\,\dota^4}
- \frac{\dddot{a}}{2\,\dota^3}
- \frac{a^2\,\dota^2}{\lp^4} \rqu
+ \vm{\hh} + \frac{\Lambda\,a^3}{6\,\lp^2}
\simeq
0
\ ,
\eeq
where $\vm{\hh}$ is given in Eq.~\eqref{hasy}.
The next step is to transform it into an equation for the Hubble parameter 
$\hub=\hub(\tau)$,
\beq
\hub^6
-2\,\lt \frac{\lp^4}{a^6} + k\,\frac{\lp^2}{a^4}\,\vm{n}+
\frac{\Lambda}{6} \rt
\hub^4
+\frac{\lp^4}{4\,a^6} \lt
2\,\hub\,\ddot{\hub}
- 6\,\hub^2\,\dot{\hub}
-5\,\dot{\hub}^2 \rt
\simeq
0
\ .
\label{eqg1}
\eeq
\par
It is worth noting that one can recover the de~Sitter solution from the above equation
for very large $a$.
In fact, if we make the ansatz $\dot\hub=0$, Eq.~\eqref{eqg1} becomes
\beq
\hub^2 = 2\,\lt \frac{\lp^4}{a^6}
+k\,\frac{\lp^2}{a^4}\,\vm{n}
+\frac{\Lambda}{6} \rt
\ ,
\label{dH0}
\eeq
and the limit $a\gg \lp$ yields the standard relation $\hub^2 = {\Lambda}/{3}$.
The second bracket in Eq.~\eqref{eqg1} therefore contains terms that qualify
as {\em non-adiabatic\/}, since they would be neglected in the
adiabatic approximation for the cosmic scale factor, that is for
$|\ddot\hub|\ll|\dot\hub|^{3/2}\ll\hub^3$~\cite{gf}.
\par
Turning again our attention to Eq.~\eqref{eqg1}, 
we recall that in Ref.~\cite{asyfree} we found that $\vm{n}\simeq 0$.
We then see that the condition~\eqref{ydef} makes the last two terms
multiplying $\hub^4$ asymptotically negligible with respect to the first one,
\beq
\frac{\lp^6}{a^6}\sim
\frac{1}{k^6}
\gg
\frac{\vm{n}}{k^3}
+\frac{1}{6}\,\lp^2\,\Lambda
\sim
\lp^2\,\hub_0^2 
\ .
\label{c_1}
\eeq
Further, the first term in Eq.~\eqref{eqg1}, which would dominate for large $a$,
can also be neglected asymptotically (with respect to the second one).
In fact, again on using Eq.~\eqref{ydef}, we obtain
\beq
a^6\,\hub^6\sim
k^6\,\hub_0^6\, \lp^6 \ll \hub_0^4\,\lp^4
\ .
\label{c_2}
\eeq
\par
Finally, Eq.~\eqref{eqg1} in the asymptotic regime can be approximated
as~\footnote{Similar equations arise in models containing both local
quadratic corrections and (in contrast to Refs.~\cite{barrow,clifton,bertolami}) 
the conformal anomaly terms.
Their solutions near the cosmological singularity were studied in
Ref.~\cite{fischetti}, while solutions with a bounded curvature and
quasi-de~Sitter behaviour were found in Ref.~\cite{starobinsky}.}
\beq
2\,\hub\,\ddot{\hub}
-6\,\hub^2\, \dot{\hub}
-5\,\dot{\hub}^2
-8\,\hub^4 
\simeq 0
\ .
\label{grav3}
\eeq
This equation (like the complete Eq.~\eqref{eqg1}) 
does not admit a constant solution for $\hub$.
A solution can instead be found in the form of a power-law,
\beq
\hub= \gamma\, \tau^\beta
\ .
\eeq
where $\gamma$ and $\beta$ are constants.
Substituting into Eq.~\eqref{grav3}, we find $\beta=-1$
and a quartic equation for $\gamma$.
Two of its four solutions coincide with $\gamma=0$, that is $\hub=0$,
which we discard, since it cannot be matched smoothly with the de~Sitter
evolution, and we are left with the solutions
\beq
\hub_{1/4}= \frac{1}{4} \, \tau^{-1}
\quad
{\rm and}
\quad
\hub_{1/2} = \frac{1}{2} \, \tau^{-1}
\ ,
\label{H1214}
\eeq
which yield the following forms for the scale factor
\beq
a_{1/4}= a_0 \lt\frac{\tau}{\tau_0}\rt^{1/4}
\ ,
\label{solgrav1_4}
\eeq
and
\beq
a_{1/2} = a_0\lt\frac{\tau}{\tau_0}\rt^{1/2}
\ .
\label{solgrav1_2}
\eeq
These expressions are, of course, understood to apply in the asymptotic regime
only, that is for $\tau\simeq\tau_0$, with $a_0=k\,\lp$.
By matching $a_{\rm dS}(\tau_0)\simeq a_0$ and~\footnote{Had we assumed
$\tau_0<0$, the Hubble parameters~\eqref{H1214} would be negative and this matching
impossible. 
The solutions~\eqref{solgrav1_4} and~\eqref{solgrav1_2} for $\tau_0<0$ might however
represent the collapse before a bounce~\cite{gf} of the form considered, for example,
in pre-Big~Bang cosmology~\cite{gasperini}.
In this context, the case $\hub=0$ could also be accepted and the asymptotic regime
interpreted as the bounce itself.}
\beq
\hub_{1/4}(\tau_0)\simeq
\hub_{1/2}(\tau_0)\simeq
\tau_0^{-1}\simeq
\hub_0
\ ,
\label{inp}
\eeq
one then finds $\alpha\sim a_0$.
Moreover, at the time $\tau=\tau_0$, the requirements~\eqref{c_1} and~\eqref{c_2}
both reduce to the same UV cut-off
\beq
k^3\ll \frac{1}{\hub_0\,\lp}
\ ,
\label{lp121}
\eeq
which is just the defining condition~\eqref{ydef} for $a=a_0$.
\par
It is important at this point to make sure that the above solutions~\eqref{solgrav1_4}
and~\eqref{solgrav1_2} are compatible with all the approximations employed.
In particular, we know that the leading order WKB approximation~\eqref{wkb}
is reliable if (see, for example, Ref.~\cite{bender})
\beq
\left|\frac{\ud P_a}{\ud a}\right|
\ll P_a^2
\ .
\label{cwkb}
\eeq
Upon substituting $a=a_{1/2}$, one finds that the corresponding momentum,
\beq
P_a=-\lp^{-2}\,a_0^2\,\hub_0
\ ,
\eeq
is constant and the condition~\eqref{cwkb} is thus trivially satisfied.
However, for $a=a_{1/4}$, 
\beq
P_a=-\frac{a_0^4\,\hub_0}{\lp^{2}\,a_{1/4}}
\eeq
and the condition~\eqref{cwkb}
becomes
\beq
y_{1/4}\ll 1
\ ,
\label{wkb1_4}
\eeq
in clear violation of Eq.~\eqref{ydef}.
This already casts serious doubts about the validity of the solution~\eqref{solgrav1_4},
but we shall further show its inconsistency in the next Section. 
\par
Let us end this Section by noting that the solution $a_{1/2}$ coincides with the evolution
of a classical universe filled with radiation, although the matter contribution was shown to
be negligible and the radiation-like behaviour is of pure gravitational origin.
Finally, the classical analogue to the solution $a_{1/4}$ would be a perfect fluid with
equation of state parameter equal to $5/3$.
\section{Reconsidering the matter equation}
\label{Smeq}
\setcounter{equation}{0}
The results presented in the previous Section assume
the asymptotic matter states obtained in Ref.~\cite{asyfree} for the
de~Sitter evolution, but lead to different expressions for the scale
factor.
It is therefore necessary to check that the new solutions~\eqref{solgrav1_4}
and~\eqref{solgrav1_2} are consistent with the chosen matter states.
\subsection{Solution $a \propto \tau^{1/4}$}
For the solution~\eqref{solgrav1_4}, one can easily see that
\beq
y_{1/4} \simeq \frac{\lp^2\,a_{1/4}}{2\,a^4_0\,\hub_0}
\ ,
\label{y14}
\eeq
which decreases for decreasing $a$, contrary to what happens in de~Sitter, and
there is no proper regime~I.
For $a_{1/4}\simeq a_0$, the right hand side of the matter equation~\eqref{mat} 
still vanishes on the asymptotic states~\eqref{wematt} and the left hand side
reduces to
\beq
\lt i\,\partial_\tau
-\hh\rt \kcs
\simeq
\lqu
i\,\partial_y
-\lt
\frac{2}{\lp^2}\,{\hat\pi}^2 +
k^2 \, \frac{a_0^{16}\,y^4}{8\,\tau_0^4\,\lp^{10}}\, \pop^2
\rt \rqu \kcs \simeq 0
\ .
\eeq
The kinetic term now dominates when $y$ is small (as also required by the
WKB approximation, see Eq.~\eqref{wkb1_4}), rather than large,
and the asymptotic matter states found in Ref.~\cite{asyfree}
seem to remain valid solutions.
However, the condition $y_{1/4} \ll 1$ estimated at the asymptotic time $\tau=\tau_0$
gives
\beq
k\gg (2\,\hub_0\, \lp)^{-1/3}
\ ,
\label{lp14}
\eeq
which violates the condition~\eqref{lp121}.
Since this implies that the conditions~\eqref{c_1} and \eqref{c_2} are also violated,
this case cannot be accepted as a valid description of the coupled dynamics.
\subsection{Solution $a \propto \tau^{1/2}$}
For the solution~\eqref{solgrav1_2}, one analogously finds that
\beq
y_{1/2}\simeq \frac{\lp^2}{2\,a^2_0\,\hub_0\,a_{1/2}}
\ ,
\eeq
which increases for decreasing $a$, like in de~Sitter. 
It then follows that, contrary to the previous case, in the regime $y_{1/2}\gg 1$
(equivalent to Eq.~\eqref{lp121} for $\tau = \tau_0$) the approximations
required for the validity of Eq.~\eqref{grav3} now hold.
At the same time, since $\Delta\hat O\ket{\chi_{\rm asy}}\simeq 0$, the
states~\eqref{wematt} remain asymptotic solutions to Eq.~\eqref{mat}.
Let us also note that the mode $k$ lies inside the Hubble radius at the asymptotic
time $\tau_0$ if
\beq
\frac{a}{k}\,\hub
\sim
\hub_0\,\lp
\ll 1
\ .
\label{c_c}
\eeq
If this condition is also satisfied, the UV cutoff on $k$ implied by Eq.~\eqref{lp121}
is large and the range of allowed momenta sizable.
\par
We can then conclude that only the solution $a_{1/2}$ in Eq.~\eqref{solgrav1_2}
can be accepted and is compatible with the asymptotic matter states obtained in
Ref.~\cite{asyfree}.
Of course, all the disclaimers discussed in that paper about the existence
of regime~I and the condition~\eqref{lp121} still apply.
It is in particular possible that the universe was born with an initial scale
factor $a\gg a_0$ and the asymptotic regime we have found never
occurred~\cite{hs}.
\section{Summary and conclusions}
\label{conc}
\setcounter{equation}{0}
We have investigated the expansion of the scale factor in the very early universe,
around the onset of inflation, by studying the gravitational equation obtained from
the Born-Oppenheimer reduction of the Wheeler-DeWitt equation for one mode of
a minimally coupled scalar field in the de~Sitter space-time.
We have found that quantum gravitational fluctuations due to the presence of matter
asymptotically vanish going ``backward in time'' toward the limit of applicability of
the semiclassical approximation, both in the equation for gravity and
that for matter.
The gravitational equation can then be solved and there emerges
a phase during which the cosmological constant and the scalar field energy
are negligible with respect to purely gravitational terms:
in a sense, gravity seems to determine its own (initial) evolution.
\par
In particular, we have uniquely obtained a ``spontaneous'' power-law evolution for
the scale factor which reproduces the usual behaviour of a radiation-filled universe.
We wish to remark once more that, similarly to the corrections near the FRW singularities
studied in Ref.~\cite{gf}, this radiation-like evolution is driven by terms in the WKB form
of the gravitational wavefunction and is thus essentially due to the initial non-classicality
of the space-time.
Inflation starts at a later stage, when the universe becomes purely
classical, and is driven by a pre-existing cosmological constant which was
ineffective during the semiclassical phase.
It is also interesting that a similar behaviour near the singularity is known to be generic
for the Einstein gravity with local quadratic corrections~\cite{barrow,clifton} and was
also analysed long ago in Ref.~\cite{bertolami} and more recently, for example,
in Refs.~\cite{sarangi,barvinsky}.
\par
Let us conclude by mentioning that our equations also suggest a different
scenario in which the asymptotic regime corresponds to a ``bounce'' of the
cosmic scale factor (that is $\hub=0$ or $\hub<0$) of the form employed in
pre-Big~Bang cosmology~\cite{gasperini}.
This possibility was however not considered in details here, since it requires
that the universe went through an intermediate stage prior to the de~Sitter
expansion for which we have no analytical description.
\section*{Acknowledgments}
C.~A.~acknowledges financial support by the University of
Bologna under the Marco Polo grants scheme and wishes to
thank the ICG, Portsmouth, UK, for their kind hospitality.
\end{document}